\documentclass[preprint2]{aastex}
\usepackage{graphics,graphicx,amsmath}

\def\ale{\mathrel{\hbox{\rlap{\hbox{\lower4pt\hbox{$\sim$}}}\hbox{$<$}}}}
\def\age{\mathrel{\hbox{\rlap{\hbox{\lower4pt\hbox{$\sim$}}}\hbox{$>$}}}}

\def\kms{{km~s$^{-1}$}}
\def\btheta{{ \mbox{\boldmath $\theta$} }}

\begin{document}

\title{100 $\mu$as Resolution VLBI Imaging of Anisotropic Interstellar Scattering towards Pulsar B0834$+$06}

\shorttitle{VLBI Imaging of Scattering Toward B0834$+$06}
\shortauthors{Brisken et al.}

\author{W.~F.~Brisken\altaffilmark{1},  J.-P.~Macquart\altaffilmark{2}}
\author{J.~J.~Gao\altaffilmark{3}, B.~J.~Rickett\altaffilmark{3},  W.~A.~Coles\altaffilmark{3}}
\author{A.~T.~Deller\altaffilmark{1}, S.~J.~Tingay\altaffilmark{2}}

\altaffiltext{1}{National Radio Astronomy Observatory, P.O.\ Box O, Socorro NM 87801, U.S.A.}
\altaffiltext{2}{Dept.\ of Applied Physics, Curtin University of Technology, Perth, WA, Australia}
\altaffiltext{3}{Department of Electrical Engineering and Computer Science, University of California, San Diego, 9500 Gilman Drive, La Jolla, CA 92093, U.S.A.}



\begin{abstract}
We have invented a novel technique to measure the radio image of a pulsar scattered by the interstellar plasma with 0.1~mas resolution.   We extend the ``secondary spectrum'' analysis of parabolic arcs by Stinebring et al.\ (2001) to very long baseline interferometry and, when the scattering is anisotropic, we are able to map the scattered brightness astrometrically with much higher resolution than the diffractive limit of the interferometer.  We employ this technique to measure an extremely anisotropic scattered image of the pulsar B0834$+$06 at 327~MHz. We find that the scattering occurs in a compact region about 420~pc from the Earth. This image has two components, both essentially linear and nearly parallel. The primary feature, which is about 16~AU long and less than 0.5~AU in width, is highly inhomogeneous on spatial scales as small as 0.05~AU. The second feature is much fainter and is displaced from the axis of the primary feature by about 9~AU. We find that the velocity of the scattering plasma is $16 \pm 10$~\kms approximately parallel to the axis of the linear feature.  
The origin of the observed anisotropy is unclear and we discuss two very different models. It could be, as has been assumed in earlier work, that the turbulence on spatial scales of  ($\sim 1000$~km) is homogeneous but anisotropic. However it may be that the turbulence on these scales is homogeneous and isotropic but the anisotropy is produced by highly elongated (filamentary) inhomogeneities of scale 0.05-16 AU.

\end{abstract}


\section{Introduction}

Radio pulsars provide a powerful tool for studying the ionized interstellar medium.  The dispersion in their pulse arrival times probes the mean electron density.  Their very small diameters ensure that they display the full range of scintillation and scattering phenomena, which probe the fine spatial structure in the electron density.  Their pulse amplitudes exhibit a combination of diffractive and refractive intensity scintillation on times from seconds to months.   Compact emission from some bright active galactic nuclei can also show scintillation on times of hours to months, albeit smoothed by the effect of their larger angular diameters.   The large body of pulsar scintillation data has been interpreted in terms of homogeneous isotropic Kolmogorov turbulence in the interstellar plasma \citep{ars95}.  See reviews by \citet{Ric90,Nar92}. 

However, a litany of observational evidence now points to the existence of compact ionized structures in the interstellar medium (ISM) whose scattering characteristics are well beyond those of such homogeneous isotropic Kolmogorov turbulence.  In particular the scattering is seldom uniformly distributed along the line of sight. It is often dominated by one local region somewhere in the line of sight, which we refer to as a ``thin screen.''  Although it is unlikely to resemble a screen, it is thin with respect to the total line of sight from the source to the observer \citep[e.g.,][]{putney,rjtr}.   Inhomogeneity in the turbulence is required on kiloparsec scales to explain how the level of pulsar scattering varies with distance and Galactic coordinates \citep{Cor91}. Inhomogeneity is also required on the parsec scale in the local ISM to explain the intermittent nature of the scintillation observed in a few quasars on hour-long time scales \citep[e.g.,][]{DTdB03,KC06}.

In addition, evidence for AU-scale inhomogeneity in the turbulence comes from extreme scattering events (ESEs), which are observed in a few quasars as rare, large (10 to 50\%) variations in flux density over several weeks \citep{Fiedler87,La01,Sen08}.  These are generally seen as a decrease in flux density attributed to the passage across the line of sight of an ionized cloud which either scatters \citep{Fiedler87} or refracts \citep{Rom87} the radiation.  

In recent years there has also been increasing evidence that interstellar scattering (ISS) is not only inhomogeneous but also anisotropic.  The most direct measure of anisotropy is through very long baseline interferometry (VLBI) imaging of scattered brightness, but it is only detectable on a few heavily scattered lines of sight \citep[e.g.,][]{Des01}.  Anisotropy in the ISS diffraction pattern has also been measured indirectly \citep[e.g.,][]{Ric02, DTdB03,Col05,Big06}.  Such observations give evidence for elongated fine structure in the ISM on scales of thousands of kilometers, suggesting anisotropic magneto-hydrodynamic turbulence controlled by the magnetic field as discussed by \citet{Gol95} and \citet{Spa99}.

The examples cited above show departures from both homogeneity and isotropy in the ionized ISM. It is possible, but by no means proven, that the various phenomena have a common origin in a population of AU-scale anisotropic regions of enhanced density and turbulence which we here generically refer to as ``clouds.''  However, we note that such localized clouds must contain fine scale substructure that causes scattering at radio frequencies.  Such a population of clouds presents a serious puzzle.  Their number density must be many orders of magnitude greater than that of stars and their implied electron densities $n_{\rm e} \age 10$~cm$^{-3}$ are much higher than expected in pressure equilibrium in the warm ionized phase ISM \citep{Rom87}.

The discovery of parabolic arcs in the ISS of pulsars by \citet{Sti01}
has added a powerful new tool for probing such clouds.  Many pulsars exhibit parabolic arcs in their secondary spectra (SS), which is the power spectrum (versus delay and Doppler frequency) of the dynamic spectrum of intensity (versus frequency and time).  The arcs are sometimes narrow (in delay) which implies scattering by a thin layer.   The distribution of power in the SS often reveals anisotropic scattering and in some cases there are discrete downwards facing ``arclets,'' which also imply scattering from isolated anisotropic clouds (``cloudlets''). See \citet{Cor06,Wal04} for interpretation of the arcs.  

The SS allows a two dimensional reconstruction of the scattered image from observations at a single receiver, since each point in the SS isolates the scintillation power associated with interference between pairs of points on the scattering disk.  However the reconstruction can be model dependent and has an inherent two-fold ambiguity \citep{Cor06,Tra07}.  Occasionally one can also see isolated peaks in the SS corresponding to narrow bandwidth fringes in the dynamic spectrum \citep{Wol87,Ric97}.  Such fringes result from interference between the normal primary (on-axis) scattering disk and the off-axis discrete cloud.   

Recent results from \citet{Hil05} have compounded the difficulties in understanding the nature of the underlying ionized clouds.  They observed the SS towards pulsar B0834$+$06 and found four distinct arclets scattered through 7 to 12~mas which they interpreted as originating from $\sim$0.2~AU clouds requiring $n_{\rm e} > 100$~cm$^{-3}$, similar to those invoked to explain ESEs towards quasars.  By monitoring the evolution of structures in the secondary spectrum, they followed these clouds over three weeks and showed that they co-moved with the rest of the scattering material.  

In this paper we report VLBI observations of the scintillation from the same pulsar (B0834$+$06) in order to further investigate these clouds.  We have developed a novel astrometry technique that makes use of SS-like quantities derived from the interferometer visibilities.  Using these ``secondary cross spectra'' (defined in Table~\ref{tab:defs}), we can accurately localize points on the scattering screen corresponding to high signal-to-noise pixels of the SS. We use the results of the astrometry of many such points to measure the distance and velocity of the interstellar clouds and so define a precise model for the scattering.  Applying this precise model to the astrometric image with the scattering model allows us to eliminate all the otherwise troublesome ambiguities and validates the model.  We then use this precise model to recover the scattered image with even greater angular resolution from the secondary spectrum itself.  In our observations the Rayleigh resolution (i.e., synthesized beam) of the VLBI array is about 35~mas, the astrometric precision is about 1~mas, and structure in the scattered image recovered by modeling is found on a scale of $100$~$\mu$as.

\section{Observations and Primary Analysis}

Pulsar B0834$+$06 was observed as part of a global VLBI project on 2005 November 12.  To obtain the greatest astrometric precision we used four of the largest telescopes in the global VLBI network: Arecibo (AO); the Green Bank Telescope (GBT); Jodrell Bank (JB); and tied-array Westerbork (WB).  We used a somewhat lower frequency than most secondary spectrum observations (327~MHz) in order to obtain higher angular scattering and thus to better resolve the image with astrometry.  Baseband data were recorded using the Mark5A disc recorders at all antennas.  Four dual circular polarization 8~MHz bands spanning the frequency range 310.5 to 342.5~MHz were recorded with four-level quantization, yielding a total data rate of 256~Mbps per antenna. In order to minimize unwanted signals that may correlate between stations, the pulse calibration signals were disabled at all of the antennas.  A total of 5700~s of on-source data were recorded.  

An initial correlation was performed using the VLBA correlator in Socorro, NM with typical continuum VLBI spectral and temporal resolutions.  These correlator products were used for delay and bandpass determination.  The raw data were recorrelated at Swinburne University with the DiFX software correlator \citep{Difx}, whose flexibility enabled the data to be processed with extremely high spectral resolution.  The output visibilities consisting of 32768 spectral channels per 8~MHz band (244~Hz resolution) were dumped every pulse period (1.25~s).  The software correlator used incoherent dedispersion to apply a 125~ms wide on-pulse bin; both the on- and off-pulse visibilities were recorded.  The timing information from which the gate ephemeris was derived was determined through simultaneous pulsar timing observations made at the GBT.

Dynamic spectra were constructed from the autocorrelation spectra generated by the correlator for each polarization.  In an analogous manner, {\em dynamic cross spectra} were formed from the cross correlation spectra (visibilities). In the case of the power spectra the off-pulse spectra were subtracted from the on-pulse spectra, removing the telescope system noise and most radio frequency interference (RFI).  This was not necessary in forming clean dynamic cross spectra because the noise and RFI are not correlated between any two antennas.  Both the dynamic spectra and the dynamic cross spectra were averaged into 5-pulse time blocks to reduce the effect of the strong pulse-to-pulse intensity fluctuations. The total power in each 8~MHz band was constrained to be constant, a reasonable condition considering that 8~MHz is much wider than the diffractive scintillation bandwidth, measured here as $\delta \nu \sim 3$~kHz, which we note is considerably narrower than 38~kHz from \citet{Cor86}. This process will suppress any refractive intensity fluctuations, however these would have a time scale of many days. 

The instrumental bandpasses were corrected using interferometric observations of a strong background source.  Delay, phase and amplitude calibration were derived from the observations of the target source itself using the coarsely averaged visibility data from the VLBA correlator.   Since this applies {\em self-calibration}, the astrometric positions which we determine below are all referenced to the centroid of the intensity.  After calibration, the two circular polarizations were summed to form the Stokes parameter $I$ in order to maximize the signal to noise ratio, giving the dynamic spectra of total intensity, $I(f,t)$, for each antenna, and visibility, $V(f,t,{\bf b})$, for each baseline ${\bf b}$. 

\subsection{Secondary Spectrum Analysis and Arcs}
\label{sec:SS}

The secondary spectra, $A(\tau,f_{\rm D})$, are squared amplitudes of the two-dimensional Fourier transforms, $\tilde I(\tau,f_{\rm D})$, of the dynamic spectra of intensity.  Here the transform variables conjugate to frequency, $f$, and time, $t$, are delay, $\tau$, and Doppler frequency, $f_{\rm D}$, respectively. The quantity $\tilde V(\tau,f_{\rm D},{\bf b})$ represents the complex two-dimensional Fourier transforms of the complex dynamic cross spectra.   As some of the quantities we discuss in this work have not yet been introduced in the literature, we summarize in Table\,\ref{tab:defs} the conventions we adopt throughout for the various products derived from the correlator auto and cross correlations.

The dynamic spectra have a spectral resolution of 244~Hz over 8~MHz and temporal resolution of 6.25~s (about 5 pulsar rotation periods) over the observation duration of 6500~s, which are well-suited to resolving the diffractive scintillation, whose time scale is $\sim1$~min and frequency scale is $\sim3$~kHz. Thus the secondary spectra have a resolution of 125~ns in delay out to a maximum of 2.05~ms. They have resolution in Doppler frequency of 0.15~mHz over a width of $\pm$~80~mHz.  Custom software was used for this and all subsequent data processing.

For the first time, dynamic spectra of right minus left circular polarization and the corresponding secondary spectra were computed to test for differential Faraday rotation in the ISM \citep{Mac00}.  No detectable signal, significant at the 0.1\% level, was found in any of these differenced spectra.  In a simple model this implies rotation measure differences across the image of less than $\sim 1.2\times10^{-3}$~rad\,m$^{-2}$ on AU scales \citep{Mac00}.

\renewcommand{\tabcolsep}{0.03in}   
\begin{figure*}[ht]
\begin{tabular}{rl}
\includegraphics[bb= 50 155 546 610,width=8.5cm,clip]{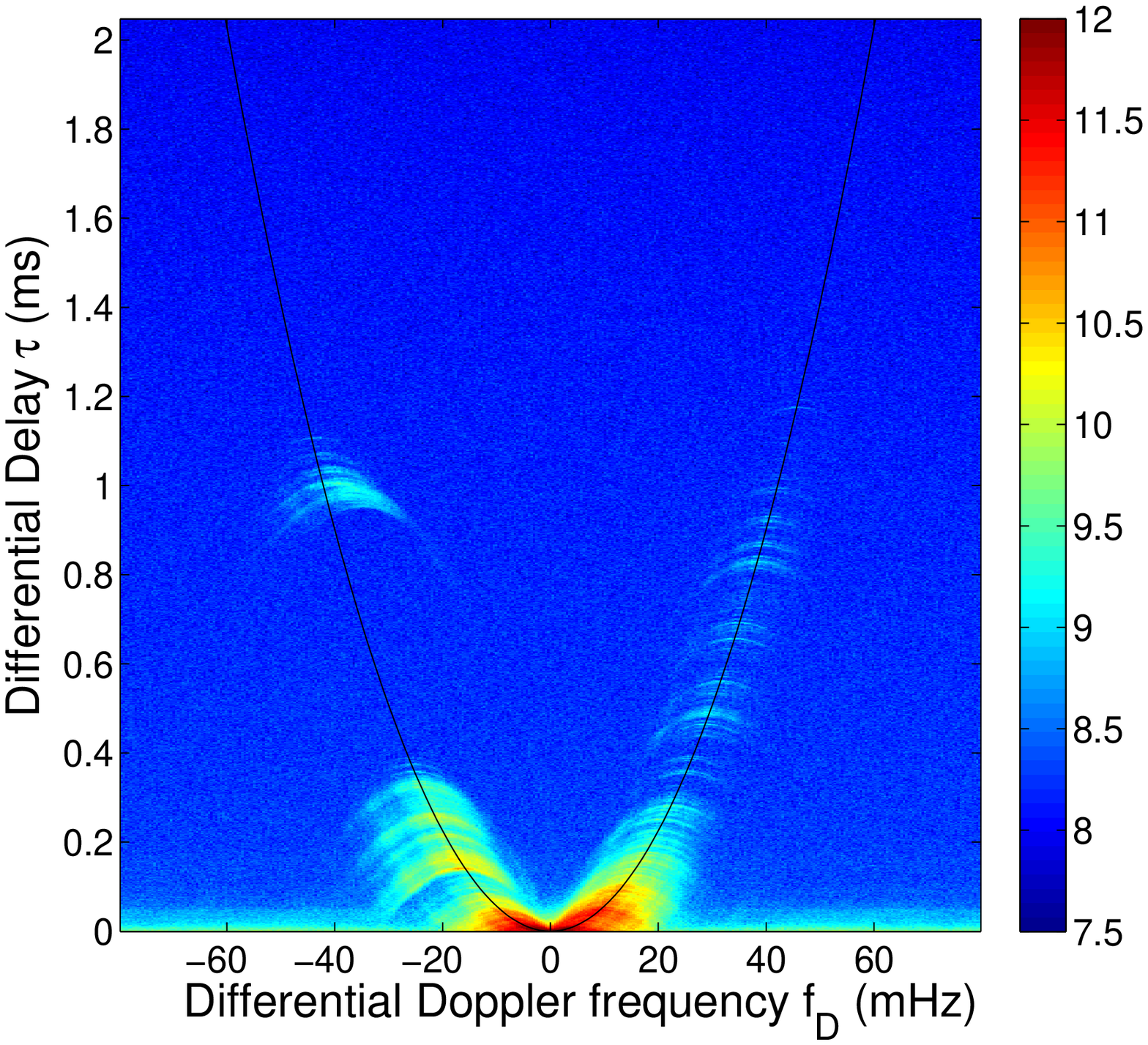} & \includegraphics[bb= 72 155 554 610,width=8.27cm,clip]{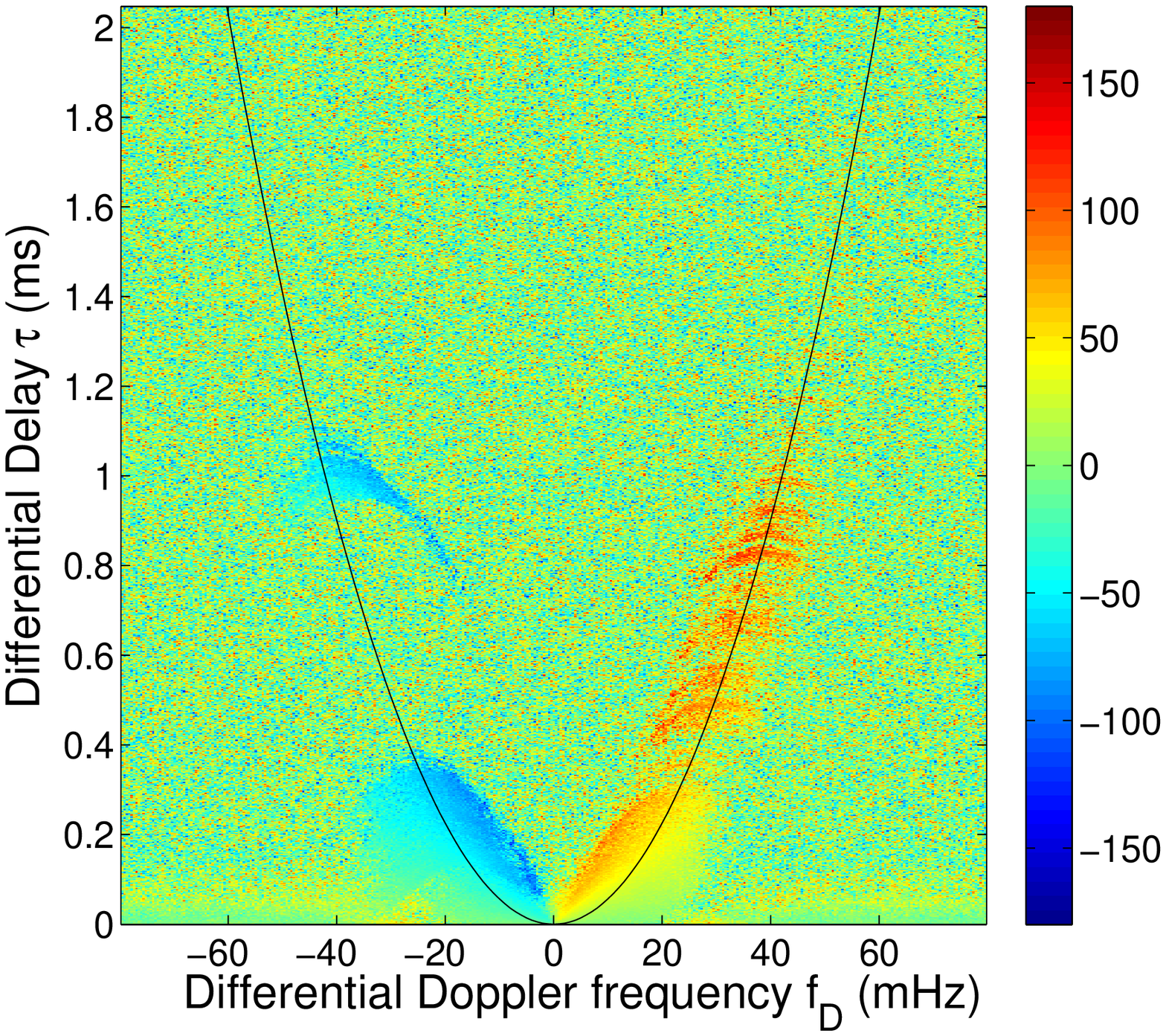} \\
\end{tabular}
\caption{The amplitude (left) and phase (right) of the seconadary cross spectrum $C$, defined in Table \ref{tab:defs}, for the 8~MHz wide 314.5~MHz sub-band on the AO-GBT baseline. Amplitude scale is $\log_{10}$ and phase is in degrees   For display purposes only, both diagrams have been smoothed over 5 pixels in delay to reduce the noise, giving plotted resolution of 0.44~mHz in Doppler frequency and $3.1$~$\mu$s in delay.   The black line is a parabola with curvature 0.56~s$^3$ on both panels. There are bright arclets in amplitude that extend along the ridge of this main parabola to delays $>1$~ms.  The phase appears much smoother, as discussed in \S\ref{sec:astrom}.
}
\label{ssObserved}
\end{figure*}
\renewcommand{\tabcolsep}{0.08in}

The amplitude of the quantity $\tilde V(\tau,f_{\rm D},{\bf b})$ $\tilde V(-\tau,-f_{\rm D},{\bf b})$ is plotted in Figure~\ref{ssObserved}.  This quantity is a generalization of the secondary spectrum to the interferometric case and will be discussed later.  The plot extends to a delay of 2~ms, which is considerably greater than has been published before, and is plotted with higher Doppler frequency and delay resolution than is usual.  It thus reveals some interesting features.  It is composed of a dense forest of fine arclets in which the apex of each arclet lies on or near the upwards facing main arc and all arcs are parabolic with the same curvature, as shown by \citet{Hil05}.
The disjoint group of arclets near 1~ms in delay with negative Doppler frequency is particularly striking and will be referred to as the 1~ms feature.  Their apexes lie inside the main arc and, as we find below, they form a separate and distinct part of the scattered image.  Some faint arclets on the positive Doppler frequency side can be identified with delays as high as 2~ms, although they are not readily visible in this figure.

\begin{table*}[ht]
{\scriptsize 
\begin{tabular}{lll}
\tableline
{\bf Term} & {\bf Symbol} & {\bf Remarks} \\
\tableline
\multicolumn{3}{l}{\bf Correlator products (2nd order in electric field)}\\
\tableline
Dynamic spectrum & $I(f,t) = \langle |E(f,t)|^2 \rangle$ & the power of the electric field as a function  \\
\null & \null & \quad of frequency and time \\
Dynamic cross spectrum & $V(f,t,{\bf b}) = \langle E_1(f,t) E_2^*(f,t) \rangle$ & the visibilities measured on a baseline ${\bf b}$ as a  \\
\null & \null & \quad function of frequency and time \\
\tableline
\multicolumn{3}{l}{\bf Intermediate quantities (2nd order in electric field)}\\
\tableline
$-$ & $\tilde{I}(\tau,f_{D}) = {\rm FT} \, [ I(f,t) ]$ & the Fourier transform of the dynamic spectrum \\
$-$ & $\tilde{V}(\tau,f_D,{\bf b}) = {\rm FT} \, [ V(f,t,{\bf b}) ]$ & the Fourier transform of the dynamic cross spectrum \\
\tableline
\multicolumn{3}{l}{\bf Derived quantities (4th order in electric field)}\\
\tableline
Secondary spectrum & $A(\tau,f_D) = |\tilde{I}(\tau,f_D)|^2$ & the squared-modulus of the Fourier transform of the dynamic \\
\null & \null & \quad spectrum \\
Secondary cross spectrum & $C(\tau,f_D,{\bf b}) = $ & the (complex) product of  $\tilde{V}(\tau,f_D,{\bf b})$ and its \\
\null & \quad $\tilde{V}(\tau,f_D,{\bf b}) \tilde{V}(-\tau,-f_D,{\bf b})$ & \quad  corresponding quantity  measured at the point $(-\tau,-f_D)$ \\
\tableline
\end{tabular}
}
\caption{Definitions of various products derived from the measured intensities and visibilities used throughout this text.  In the definitions for the first two quantities, $E$ is the electric field as measured by the antenna receiver and the angle brackets denote a time average over the measurement interval much longer than the inverse bandwidth.} 
\label{tab:defs}
\end{table*}

\section{Theory of Secondary Cross Spectra}
\label{sec:theory}

Here we briefly outline the theory of the arcs and refer the reader to \citet{Wal04} and \citet{Cor06} 
for more complete analyses.  Points distributed along the main parabolic arc are due to interference between a highly scattered component at $\btheta$ and the components arising from near the pulsar brightness centroid at $\btheta = 0$. In this case there is a differential delay $\tau = \theta^2 D_{\rm eff}/2c$ and differential Doppler frequency\footnote{Henceforth these terms will be abbreviated to delay and Doppler.} $f_{\rm D} = {\bf V}_{\rm eff} \cdot \btheta /\lambda$. Here $D_{\rm eff}$ and ${\bf V}_{\rm eff}$ are the effective distance and velocity vector of the scattering screen as defined in \S\ref{sec:scatt_model}, and $\lambda$ is the observed wavelength.  If the scattered image is essentially linear, through the origin at angle $\alpha$ with respect to ${\bf V}_{\rm eff}$, we can eliminate $\theta$ and obtain
\begin{equation}
\tau = a f_{\rm D}^2\,, \mbox{ where}\,\, a= D_{\rm eff} \lambda^2/(2 c V_{\rm eff}^2 \cos^2\alpha)
\end{equation}
as the equation of the main parabola.  Note that $a \propto \lambda^2$ regardless of the wavelength scaling of scattering angle $\theta$, as confirmed experimentally by \citet{Hil03}.

To understand the interferometric observations consider the pulsar as a point source at a distance $D_{\rm p}$ from Earth, whose radiation is scattered by an inhomogeneous region of ionized interstellar medium at a distance $D_{\rm s}$ from Earth.  In this thin-screen approximation the phasor due to radiation scattered at a point ${\bf x}_j= D_{\rm s} \btheta_j$ in the screen suffers a phase delay $\phi_j$ due to the scattering plasma 
and is received with a total phase delay $\Phi_j({\bf r}) = \phi_j + k({\bf x}_j-\beta {\bf r})^2/2 \beta D_{\rm s}$ where $\beta=1-D_{\rm s}/D_{\rm p}$ and ${\bf r}$ is the transverse position of the receiving antenna.  The received electric field is the summation of all such scattered components, either expressed by the Fresnel diffraction integral \citep[see][]{Cor06} or approximated by a summation over all stationary phase points \citep[see][]{Wal04}, in strong scattering.  

Here we use the stationary phase approximation to obtain the correlation of the fields between antennas at locations $-{\bf b}/2$ and ${\bf b}/2$.  This defines the visibility whose Fourier transform in time and frequency can be written as:
\begin{eqnarray}
\tilde V(\tau,f_{\rm D},{\bf b}) &=& \sum_{j,k} \exp[i (\Phi_j({\bf -b}/2)-\Phi_k({\bf b}/2))] \nonumber \\ 
&\times& \mu_j \mu_k \delta(f_{\rm D}-f_{{\rm D},jk}) \delta(\tau-\tau_{jk}). \label{eq:VisSS}
\end{eqnarray}
where:
\begin{subequations}
\begin{eqnarray}
f_{{\rm D},jk} &=&  \frac{1}{\lambda} (\btheta_j -\btheta_k) \cdot {\bf V}_{\rm eff}, \label{DopplerRate}\\
\tau_{jk} &=& \frac{D_{\rm s}}{2\,c\,\beta} (\theta_j^2 - \theta_k^2) + \left[ \frac{\phi_j}{2\pi \nu} - \frac{\phi_k}{2\pi\nu} \right] . \label{delay} 
\end{eqnarray} \label{delayrate}
\end{subequations}
The $\delta$ functions in equation~(\ref{eq:VisSS}) determine how each particular point on the secondary spectrum is related to a pair of points in the screen.  With finite bandwidth and observing time the $\delta$ functions should be replaced by finite narrow sinc functions.  The equation sums over all possible pairs of stationary phase points in the screen, with $\mu_j,\mu_k$ as the magnifications determined by phase curvature of each point \citep[see][]{Wal04}.  In the full Fresnel formulation, the summation becomes a double two-dimensional integration over the screen with all $\mu=1$.   The plasma delay term in square brackets in equation~(\ref{delay}) involving $\phi_j$ is unimportant compared to the first term when constructing secondary spectra in strong scintillation, but in any case this term cancels in the astrometry discussed below.   

In either formulation equations (\ref{DopplerRate}) and (\ref{delay}) connect the points on the scattering screen to $\tau,f_{\rm D}$ in the secondary spectrum and are independent of ${\bf r}$. Note that $f_{\rm D}$ can be understood as the differential Doppler shift between two waves scattered at differing points on the screen \citep{Cor06} and that the delay $\tau$ is the differential group delay evaluated at the center of the observing band.  

Each arclet arises due to the interference of radiation from a given fixed point on the scattering screen with scattered radiation from $\btheta \sim 0$. Examination of the delay in equation (\ref{delay}) shows that for a strong contribution from fixed $\theta_j$ the greatest delay (i.e., arclet apex) will be where $\theta_k$ goes through zero.  The particular shape of each of the arclets reflects the scattered brightness distribution near core of the image.  Furthermore the fact that the SS amplitude is bright over a {\em narrow} range in delay shows that the points comprising the bright emission centroid region are extended along a line rather than in a circular halo \citep[see][]{Cor06,Wal04}.

The velocity of the scintillation pattern across the line of sight is ${\bf V}_{\rm ISS} = (1-\beta) {\bf V}_{\rm p} + \beta {\bf V}_{\earth} - {\bf V}_{\rm s}$, where ${\bf V}_{\rm s}$, ${\bf V}_{\earth}$ and ${\bf V}_{\rm p}$ are the velocities of the screen, Earth and pulsar respectively \citep[see e.g.,][]{Cor98}.  The scintillation velocity is dominated by the pulsar, whose proper motion \citep{Lyn82} combined with a distance of 643~pc (estimated from the dispersion measure using the NE2001 Galactic electron distribution model of \citet{NE2001}) gives $V_{{\rm p},\alpha} = 6.1_{-15}^{+15}$~\kms, $V_{{\rm p},\delta} = 156_{-18}^{+15}$~\kms; note that these velocity errors do not include the substantial uncertainties in the pulsar distance.  

\subsection{Astrometric Imaging}
\label{sec:astrom}

In single dish observations information on the scattered image is limited to the power in the secondary spectrum, obtained from $\tilde I(\tau,f_{\rm D})$.  However, the addition of interferometric observations permits high-precision astrometry to be performed on each component of the secondary cross spectrum, which isolates the wavefield due to a single pair of interfering waves.  By applying astrometry to each such pair individually, we can recover an apparent image of the scattered pulsar radiation.  

For a particular pixel from $\tilde V(\tau,f_{\rm D},{\bf b})$ equation (\ref{eq:VisSS}) sums over all pairs of points that satisfy equations (\ref{delayrate}).  However, we now consider the case that only one pair of points satisfies the latter conditions, a condition that is valid for a linear or sufficiently anisotropic source.  Then the phase of $\tilde V$ is 
$\Phi_{j,k}=\Phi_j({\bf -b}/2) - \Phi_k({\bf b}/2)$ is
\begin{equation}
\Phi_{j,k}= \phi_j-\phi_k + \frac{k D_{\rm s}}{2\beta} [ \theta_j^2-\theta_k^2 +\frac{\beta {\bf b}}{D_{\rm s}}\cdot (\btheta_j+\btheta_k) ]
\end{equation}
which has terms that are symmetric and anti-symmetric in scattered points $j,k$.  
We cancel the anti-symmetric terms by summing the phase of $\tilde V$ at each point $(f_{\rm D}, \tau)$ and its conjugate point at $(-f_{\rm D}, -\tau)$. This eliminates the random screen phase leaving the symmetric part of the phase as
\begin{equation}
\psi_{j,k}({\bf b}) = \Phi_{j,k}+\Phi_{k,j} = \frac{2\pi}{\lambda} {\bf b} \cdot ({\btheta}_j + {\btheta}_k) \, ,
\label{psi}
\end{equation}
which neatly encodes the average position of the two scattering points projected parallel to the baseline ${\bf b}$.  As we show below, the scattering is indeed highly anisotropic and the assumption of a single pair of interfering waves for each pixel is well justified.

In practice the anti-symmetric phases are cancelled by first forming the secondary cross spectrum, defined as
\begin{equation}
C(f_{\rm D}, \tau,{\bf b}) = \tilde V(f_{\rm D}, \tau,{\bf b}) \tilde V(-f_{\rm D}, -\tau,{\bf b}) \, ,
\label{scs}
\end{equation}
for all non-negative values of $\tau$.  
Subsequently, the resultant complex phasors are averaged over small regions (3 pixels in Doppler by 5 in delay) in order to reduce the effect of noise before the argument is determined: 
\begin{equation} 
\psi({\bf b}) = \mbox{arg}\left(\langle  C(f_{\rm D}, \tau,{\bf b}) \rangle \right) \, .
\end{equation}
Here $\langle \rangle$ denotes the 3 by 5 averaging which properly weights the complex products before the phase is computed.  The result is the geometric phase of the secondary cross spectrum sampled with a resolution of $0.63$~$\mu$s in delay and 0.44~mHz in Doppler.  As with any phase measurement the result is modulo $2\pi$.


The right panel of Figure~\ref{ssObserved} shows $\psi$, the phase of the secondary cross spectrum, for the baseline from GBT to AO.  In contrast to the sharply defined arclets in the amplitude plot, the phase is relatively smooth, appearing continuous between the arclets, as might be expected if the fine structure in amplitude comes from a continuous image with fine structure in brightness.   Consequently we can use astrometry from $\psi$ to actually map out the scattered brightness distribution with remarkable precision.  Furthermore we can measure the position angle of the axis of elongation and estimate both the distance to the region of scattering and its velocity.

We can use equation~(\ref{psi}) to find  $(\btheta_j + \btheta_k)$ for any point in the secondary cross spectrum if it is dominated by a {\em single pair}  ($\btheta_j,\btheta_k$) and there are at least two baselines with sufficient signal-to-noise ratio (S/N).  However if we can identify the apex of an arclet we know $\btheta_j$ or $\btheta_k$ is zero, so we have a unique solution for the remaining angle.  The amplitude plot in Figure~\ref{ssObserved} shows that the apexes of the identifiable arclets lie near the parabolic curve shown as a line.  Hence we sampled the secondary cross spectrum along that curve to approximate a set of apex positions even where the arclets are too densely spaced to be identified individually. We defined the main parabolic arc by the curvature  $a = 5.577\times 10^{16}/f^2$ (s$^3$) for center frequency $f$~Hz in each of our four sub-bands, as shown by the black line in Figure~\ref{ssObserved}.  

\begin{figure}
\centerline{\includegraphics[width=10cm]{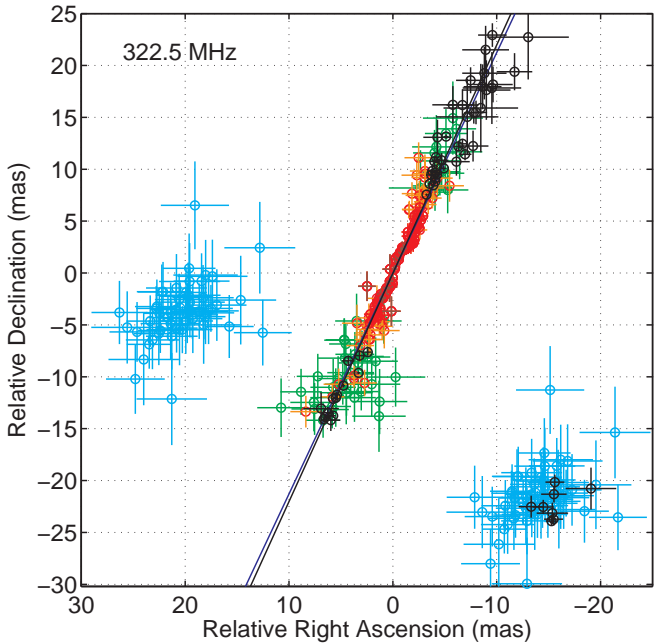}}
\caption{Map of the astrometric positions of the scattered radiation from samples along the main parabola in the sub-band at 322.5~MHz plotted as error bars in colors of green, orange, red \& dark red, for amplitude, $|C|$, increasing by factors of 10.  Right ascension and declination values are relative to the centroid position as determined by VLBI self-calibration.  Note the increase of astrometric errors with decreasing amplitude $|C|$.  The blue straight line is an unweighted fit to these points.  The black error bars are from the apexes of identified arclets and the black line is a fit to these points.   Samples from the 1~ms feature (negative Doppler and delay $> 0.9$~ms) are shown by cyan points for two possible lobe choices. As discussed in \S\ref{sec:scatt_model} the correct lobe position is at the lower right.  This is overplotted with black error bars from the apexes of individual identified arclets near 1~ms.}
\label{AstrometryFig}
\end{figure}

We selected data with a high enough S/N to estimate $\btheta$ as follows.
For each of the six baselines we found the phasor $\langle C(f_{\rm D}, \tau,{\bf b}) \rangle$ and defined $S$ as the S/N relative to the root mean square value of $| \langle C \rangle |$ in a low power (noise dominated) region near the point being examined.  Samples with delays within $\pm 0.6$~$\mu$s of the main parabola were selected if $S>4$  on the GBT-AO baseline and if $S>3$ on the WB-AO baseline.   

For each sample of $\langle C \rangle$ we first formed a  ``dirty image'' from the 6 baselines in the same sense as for traditional synthesis imaging \citep{ASP180}.  At the dirty image maximum a two-dimensional Gaussian was fit to determine a position, and both $S$ and the shape of the Gaussian were used to assign positional uncertainties. However, since $C$ is a product of two visibilities this method gave too much relative weight to the high S/N baselines and led to apparent positional errors much smaller than the scatter of the points from neighbouring samples along the main arc.  In addition there are lobe ambiguities in the synthesized beam, since two of four antennas (JB and WB) are relatively close to each other and so probe nearly the same spatial frequencies.  The result is a dirty beam that is nearly doubly periodic with lattice vectors ${\bf a}_1 = (59.0, -61.9)$~mas and ${\bf a}_2 = (34.2, 17.8)$~mas at 322.5~MHz (scaling linearly with wavelength).

After some investigation we adopted a simpler astrometric method based on a weighted linear least squares fit of the observed phases, $\psi$, using equation (\ref{psi}) (with one of the angles set to zero).  Under conditions of high S/N one can show that the phase error is $\propto 1/\sqrt{S}$ and so we used weights $\propto \sqrt{S}$.  Because the phase is observed modulo $2\pi$ there are ambiguities in position, corresponding to the lobe ambiguities. Consequently, we used a first guess position as an initial model and wrapped the phase (e.g., on the transatlantic baselines) to be within $\pm \pi$ of that solution before applying the least squares fit.  For the points from the main arc our first guess was at the origin and the results are plotted as error bars (in a pale color) in Figure~\ref{AstrometryFig}.  Points from the main arc form a continuous elongated image close enough to the origin that the phase ambiguity is not a problem for them. 

An empirically determined line passing through arclet apexes was used to sample the points in the SS associated with the 1~ms feature. The astrometry of these points, however, forms a distinct group offset from the origin in celestial coordinates, making it difficult to resolve the phase ambiguities.  Accordingly, we used first guess positions from the two closest synthesized beam lobes which are roughly equidistant from the origin.  The cyan points on the left show the fitted positions starting from the central lobe; the cyan points on the right show them starting from a position lobe-shifted by the lattice vector $-{\bf a}_2$, which corresponds to the addition of $2\pi$ to the phase on the transatlantic baselines.  In \S\ref{sec:scatt_model} we discuss the resolution of this ambiguity in favor of the right hand position.  

The major conclusion from Figure~\ref{AstrometryFig} is that the scattered image from the main arc is remarkably extended along a straight line with very few points more than $2\sigma$ from that line.  As noted above highly anisotropic scattering was already inferred by the emptiness of the secondary spectrum (particularly the absense of power at $f_D \sim 0$, $\tau > 0$) and the many discrete arclets. The apparent flaring of the points away from the origin is due to the increasing size of the error bars because the SS amplitude decreases with angle of scattering.  We define the straight line fitted to the points as the ``scattering axis''.  Note that the plot is not a proper image of the scattered brightness, since the amplitude $|C|$ must be transformed by a Jacobian into the brightness.  Hence the axial ratio is not easily estimated and we postpone the discussion of it to a later section.

\subsection{Arclet Apex Astrometry}
\label{sec:apxastrom}

The astrometric positions discussed in the preceding section and plotted in Figure~\ref{AstrometryFig} are based on the assumption that samples along the main arc correspond to the apexes of an underlying arclet.  However this is not always true, as is obvious in the 1~ms feature. So we have made an effort to identify all the discrete arclets, estimate the apex of each in an optimal way, and then map the phases at each apex to an angular position.  At 322.5~MHz we found 62 well defined arclets (with apex delays $\tau > 0.1$~ms) of which 7 are in the 1~ms feature.  At smaller delays the arclets are too crowded to identify separately.  

We formed a template by averaging all the arclet amplitudes, centered by their apex, and we fitted a straight line in phase versus $f_{\rm D}$ to the template to obtain the phase gradient versus $f_{\rm D}$ in the template. We then fitted the amplitude and phase templates to each arclet using a weighted least squares fit.  Hence we obtained estimates for the delay, Doppler, phase, and amplitude at the apex and errors in all four parameters. We then estimated $\btheta$ and its error at each apex and repeated the procedure for all 4 sub-bands.

The astrometric positions and errors of the 62 arclets are plotted in black in Figure~\ref{AstrometryFig}. The error bars are smaller than from the main arc because the phases were determined from a weighted average of the phases over the entire arclet. In addition the scatter should be smaller because they are known to correspond to interference with the bright centroid component at the origin.  However, the lobe uncertainty in the location of the 7 arclets from the 1~ms feature remains.  As we discuss below, we resolve the ambiguity in favor of the lower right lobe position marked by the black error bars derived from the 7 apexes.  The apexes near the main arc have positions extending out to 20~mas from the origin and show good linear alignment with the inner points from the main arc. A straight line fitted to the inner main arc has a position angle of $-25.2 \pm 0.5^{\circ}$ (east of the declination axis)\footnote{Here and elsewhere in this paper, reported uncertainties are 68\% confidence intervals.}.  We use this axis to define $\theta_{\parallel}$ and define $\theta_{\perp}$ by a $90^{\circ}$ clockwise rotation from it.  

We have used three techniques to resolve the lobe ambiguity.  The first relies on the frequency dependence of the positions obtained from each arclet apex.  As discussed below in \S\ref{sec:scaling} we find that the angular positions of the scattered features that cause the arclets are essentially independent of frequency.  Thus we can use the fact that the correct lobe choice should give a position that is independent of frequency.  We computed the 2 dimensional mean for the 1~ms apexes from a weighted average position of its imaged points separately for each of the four sub-bands.  For the lobe position at a negative RA offset in Figure~\ref{AstrometryFig} the mean position was independent of frequency within the errors of estimation ($\sim 0.1$~mas).  
All other lobe choices show systematic frequency dependence in their positions in agreement with that expected for a lobe error.

Confirmation comes from two further tests. The second technique relies on equation (\ref{delay}) relating delay to the angular offset of an arclet apex and is discussed in \S\ref{sec:distest}.   Similarly, the third technique relies on equation (\ref{DopplerRate}) for the Doppler frequency and is discussed in \S\ref{sec:Vest}.

\subsection{Frequency Scaling}
\label{sec:scaling}

\citet{Hil05} found the wavelength scaling of the apex Doppler frequencies of the four arclets which they identified in pulsar B0834$+$06 to scale $\propto \lambda^{-1}$.  Hence using equation (\ref{DopplerRate}) they concluded that the scattering angle responsible for the arclet was independent of wavelength.  We have done a similar analysis on the apex Doppler frequencies of all the identified arclets in our four sub-bands.  In addition we analyzed the wavelength scaling of the apex delay in the four sub-bands, which is $\propto \theta^2$ as in equation (\ref{delay}).  We combined the fits to estimate a single scaling parameter $\gamma$ where $\theta \propto \lambda^\gamma$. We estimated $\gamma$ separately for three groups of apexes and tabulate the values in Table~\ref{tab:scaling}.

\begin{table}[ht]
\caption{Wavelength-scaling exponents, $\gamma$, in the scattering angles ($\theta \propto \lambda^\gamma$) estimated from apex positions for three groups of arclets.}
\begin{tabular}{ll}
\tableline
Arclet apex group & $\gamma$  \\
\tableline
$\tau\sim 1$~ms $f_{\rm D}<0$ & $0.062\pm 0.006$ \\
$0.1\la\tau\la0.4$~ms & $0.01 \pm 0.01$\\ 
$\tau>0.4$~ms $f_{\rm D}>0$ & $0.019 \pm 0.004$\\
\tableline
\end{tabular}
\label{tab:scaling}
\end{table}


The results show that the scattering responsible for each arclet originates at a location that is essentially independent of wavelength across the 10\% range spanned by the four sub-bands.  This is entirely incompatible with the value $\gamma = 2$, to be expected if the individual arclets came from separate ray paths caused by refraction due to differing gradients in the column density of electrons.  The conclusion applies to all arclets in the SS, with the possible exception for the 1~ms feature whose position scales weakly $\lambda^{0.06}$. Such a weak scaling might apply if the arclet were caused by a lens-like concentration of plasma, such as that invoked by \citep{Rom87} to explain ESEs, but with a transverse extent very much smaller than than its distance from the direct path.

\section{Scattering Model}
\label{sec:scatt_model}

We now develop a model for the distance and velocity for the various features in the scattered image.   

\begin{figure}[ht]
\centerline{\includegraphics[width=9cm]{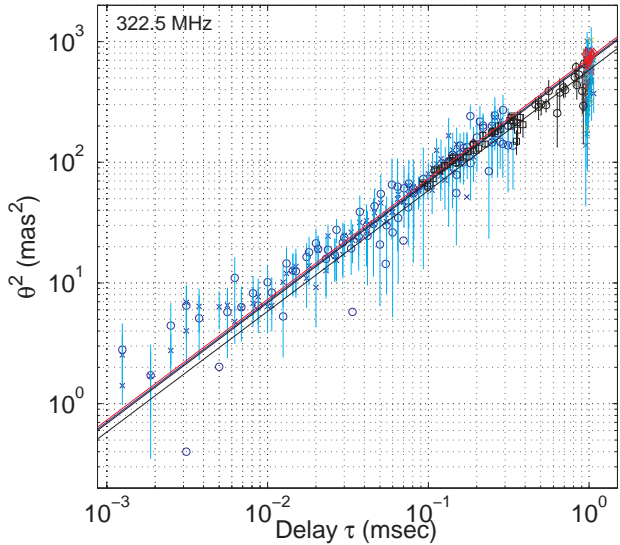}} 
\caption{$\theta^2$ versus delay.  Points from the main arc in Figure~\ref{AstrometryFig} are shown as blue $\times$ ($f_{\rm D}>0$) and $\circ$ ($f_{\rm D}<0$); the apexes of the identified arclets are plotted in black, except for those from the 1~ms feature which are in red.  The straight line models ($\theta^2 = m \tau$; see equation(\ref{eq:tau_apex})) were fitted separately to various subsets of the points to estimate their distances.
}
\label{Zeff}
\end{figure}

\subsection{Scattering Distance Estimation}
\label{sec:distest}

Equation~(\ref{delay}) relates delay $\tau$ to the angular offset $\theta$ for an arclet apex, for which the second angular component is at the origin. Thus:
\begin{equation}
\tau = (D_{\rm eff}/2c) \theta^2
\label{eq:tau_apex}
\end{equation}
where
\begin{equation}
D_{\rm eff}=D_{\rm p} (1-\beta)/\beta
\label{eq:Deffbeta}
\end{equation}
In Figure~\ref{Zeff} we plot on logarithmic scales $\theta^2$ from the astrometry against delay with the various groups of data from Figure~\ref{AstrometryFig}. Note that there is a bias to $\theta^2$ due to the uncertainties in right ascension and declination. Since we have estimates of those uncertainties we subtracted the bias before fitting. 

\begin{deluxetable}{lllllll}
\tabletypesize{\scriptsize}
\tablecolumns{7}
\tablewidth{0cm}
\tablecaption{Effective Distances and Velocities from the VLBI astrometry
\label{tab:DeffVeff}}
\tablehead{
\cutinhead{Main Parabolic Arc}
\colhead{Frequency}    &  \multicolumn{3}{c}{$D_{\rm eff}$ (pc)} & \colhead{}    &  \multicolumn{2}{c}{$V_{\rm eff}$ (\kms)} \\
\cline{2-4} \cline{6-7} \\
\colhead{MHz} & Inner Arc & \colhead{}   & $\tau\sim 1$~ms & \colhead{}    & $V_{\rm eff \parallel}$\tablenotemark{a} & $V_{\rm eff \perp}$\tablenotemark{b} \\
\cline{1-7} \\
314.5 &  1197 $\pm$ 23 & \nodata &  1350 $\pm$ 65 &  \nodata & 303.3 $\pm$  3.1 &  -131.9 $\pm$  6.4 \\
322.5 &  1175 $\pm$ 24 & \nodata & 1272 $\pm$ 29 &  \nodata & 300.7 $\pm$  3.3 &  -151.5 $\pm$ 11 \\ 
330.5 &  1195 $\pm$ 29 & \nodata & 1294 $\pm$ 20 &  \nodata & 303.3 $\pm$  3.7 &  -161.5 $\pm$ 16 \\ 
338.5 &  1133 $\pm$ 18 & \nodata & 1287 $\pm$ 24 &  \nodata & 300.0 $\pm$  2.5 &  -133.8 $\pm$  5.0 \\ [0.03in]
\cline{1-7} \\
326.5\tablenotemark{c} &  1175 $\pm$ 29 & \nodata & 1301 $\pm$ 40 &  \nodata & 301.8 $\pm$  3.2 &  -144.7 $\pm$  12 \\ 
\cutinhead{Apexes of Identified Arclets}
\colhead{Frequency}    &  \multicolumn{3}{c}{$D_{\rm eff}$ (pc)} & \colhead{}    &  \multicolumn{2}{c}{$V_{\rm eff}$ (\kms)}  \\
\cline{2-4} \cline{6-7}\\
\colhead{MHz} & $0.1\la\tau\la0.4$~ms & $\tau>0.4$~ms $f_{\rm D}>0$ & $\tau\sim 1$~ms & \colhead{}    & $V_{\rm eff \parallel}$\tablenotemark{a} & $V_{\rm eff \perp}$\tablenotemark{b}  }
\startdata
314.5 &  1165 $\pm$  19 &  1370 $\pm$  43 &  1064 $\pm$  63 &  \nodata & 305.9 $\pm$  2.9 &  -136.1 $\pm$  4.4  \\ 
322.5 &  1205 $\pm$  19 &  1419 $\pm$  67 &  1132 $\pm$  24 &  \nodata & 313.4 $\pm$  2.5 &  -152.9 $\pm$  3.7 \\ 
330.5 &  1162 $\pm$  17 &  1406 $\pm$  39 &  1180 $\pm$ 110 &  \nodata & 306.8 $\pm$  2.3 &  -158.4 $\pm$ 11 \\ 
338.5 &  1139 $\pm$  15 &  1315 $\pm$  67 &  1107 $\pm$  17 &  \nodata & 305.4 $\pm$  2.4 &  -134.9 $\pm$  1.8 \\ [0.03in]
\cline{1-7}
326.5\tablenotemark{c} &  1168 $\pm$ 23 &  1378 $\pm$  60 &  1121 $\pm$ 59 &  \nodata & 307.9 $\pm$  3.3 &  -145.6 $\pm$  8.6 \\ 
\enddata
\tablenotetext{a}{Estimates of  $V_{\rm eff\parallel}$ were derived from the main arc, since they lie along the scattering axis $\theta_{\parallel}$}
\tablenotetext{b}{Estimates of  $V_{\rm eff\perp}$ were derived from the 1~ms points, by assuming that they have the same $V_{\rm eff\parallel}$ as for the main arc}
\tablenotetext{c}{Averages from all four sub-bands with errors that include the uncertainties in each sub-band. }
\end{deluxetable}

Two separate straight lines through the origin were fitted to the (blue) points sampled along the main parabola and from the 1~ms feature.  The apex points were also fitted separately for three regions: for apex delays less than 0.4~ms, delays bigger than 0.4~ms with positive Doppler and with negative Doppler (i.e., the 1~ms feature).  The fits were weighted proportional to the reciprocal of the standard deviation in $\theta^2$. The derived effective distances are given in Table~\ref{tab:DeffVeff} with their formal 1$\sigma$ errors for all 4 sub-bands. 

The simplest model puts all of the scattering in a single relatively thin region at the same effective distance.  Comparing $D_{\rm eff}$ for the differing regions of the SS, we find $D_{\rm eff}$ in the range 1100 to 1400~pc.   Since we found no significant scaling  of the apex $\theta$ with frequency (see \S\ref{sec:scaling}), we combine data from all four sub-bands to create a combined estimate shown as a fifth row in the table.   We note that there is a possible contribution to inconsistencies among the sub-bands from a partial narrowing of the two outer sub-bands at AO by a front-end bandpass filter.

The average of the estimates in column 2 from delays less than 0.4~ms from both the main arc and from the arclet apexes is $1171 \pm 23$~pc.  In column 4 the 1~ms cluster gives $1301 \pm 40$~pc, which overlaps at $\pm 2 \sigma$ with the value $1121\pm 59$~pc obtained from the 1~ms apexes.  This small discrepancy appears in Figure~\ref{AstrometryFig} as an offset in the centroid of the blue and black points. Since the arclet apexes have been specifically identified so that one of the $\btheta$ parameters can confidently be set to zero and since $D_{\rm eff}$ for the 1~ms apexes is within $1\sigma$ of those from the main arc, we conclude that the scattering region responsible the 1~ms feature is located at the same distance as that causing the main arc.   Without the lobe-shift a fit for $D_{\rm eff}$ gives values from 1700 to 2600~pc. The large discrepancy in this distance provides the second confirmation of the lobe-shift for the 1~ms points. 

A possibly significant difference is seen between the apexes with positive Doppler and delays $> 0.4$~ms in column 3, which give a significantly larger $D_{\rm eff} = 1378\pm 60$~pc.  Whereas all of the sub-bands give a larger distance there are only a few identified arclets in each sub-band and they are of relatively low S/N in all 4 sub-bands so that accurate correction for the noise bias to $\theta^2$ becomes more critical.  Thus the evidence is relatively weak for a different distance for these points at large delay and positive Doppler.

We now take $D_{\rm eff} = 1171$~pc, combined with the pulsar distance of 640~pc in equation (\ref{eq:Deffbeta}) to obtain the fractional scattering distance $\beta = 0.353 \pm 0.005$. This is consistent within $2\sigma$ of the value of $0.29 \pm 0.04$ obtained by \citet{Hil05} using the same pulsar distance but without the advantage of knowing the full geometry of the scattering screen.  However, its true uncertainty is dominated by uncertainty in the pulsar distance which is estimated from the dispersion measure.  Dispersion measure distances are notoriously unreliable, especially for nearby pulsars and thus are only accurate to $\sim 40$\% \citep{B02}; this gives a screen distance range of 250 to 580~pc from the Earth. 

\subsection{Velocity Estimation}
\label{sec:Vest}

\begin{figure}[ht]
\centerline{\includegraphics[width=9cm]{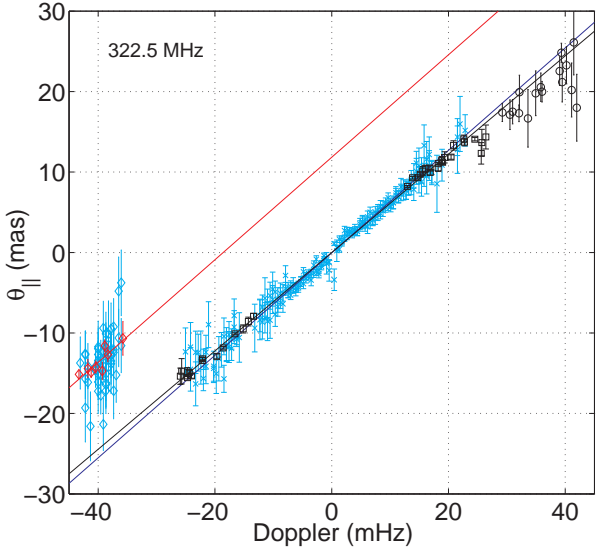}} 
\caption{$\theta_{\parallel}$  plotted against the Doppler $f_{\rm D}$ for the sub-band at 322.5~MHz.  Points plotted with cyan $\times$ marks are sampled from the inner main arc in Figure~\ref{AstrometryFig} and from a line through the 1~ms feature. Black points are from apexes of identified arclets, except 1~ms apexes which are red. 
}
\label{fig:Vest}
\end{figure}

An independent analysis is provided by equation (\ref{DopplerRate}), relating the Doppler frequency for the arclet apexes to their astrometric position.  At each apex the Doppler frequency can be written as:
\begin{eqnarray}
f_{\rm D} &=& \btheta\cdot {\bf V}_{\rm eff}/\lambda \nonumber \\
    &=& (\theta_{\parallel} V_{\rm eff\parallel} + \theta_{\perp} 
        V_{\rm eff\perp})/\lambda.
\label{eq:fDapex}
\end{eqnarray}
where
\begin{equation}
{\bf V}_{\rm eff} = {\bf V}_{\rm ISS}/\beta \approx {\bf V_{\rm p}}(1-\beta)/\beta .
\label{eq:Veff}
\end{equation}
The final approximation ignores the velocities of the Earth and of the screen relative to that of the pulsar. Then the ratio $V_{\rm eff}/D_{\rm eff}$ estimates the pulsar proper motion independent of its distance and of $\beta$.  However, here we do not make that approximation but use the measured proper motion to estimate the velocity of the scattering screen. 

In Figure~\ref{fig:Vest} we plot $\theta_{\parallel}$ against $f_{\rm D}$ which would give a straight line whose slope gives an estimate of $\lambda/V_{\rm eff\parallel}$, providing that the value of $\theta_{\perp} V_{\rm eff\perp}$ were the same or negligible for all points.  As shown in Figure~\ref{AstrometryFig} the points fitted by the straight line have small $\theta_{\perp}$ scattered about zero.  Hence we were able to estimate $V_{\rm eff\parallel}$ from the various groups of points and tabulate the results in Table~\ref{tab:DeffVeff} excluding all the points with delays $>0.4$~ms.  

If we combine inner apex fits with values from the main arc we obtain $V_{\rm eff\parallel} =305\pm3$~\kms.   We now assume the same $V_{\rm eff\parallel}$ to apply to the 1~ms apexes and since these points have a significant measured $\theta_{\perp}$ we can fit for the perpendicular velocity also, obtaining $V_{\rm eff\perp} = -145\pm9$~\kms.

In summary the scattering model provides three observable parameters which we list in Table~\ref{tab:params}.  We can express the velocities in RA, Dec and relate them to the measured pulsar proper motion velocity  $V_{{\rm p},\alpha} = 6.1$~\kms, $V_{{\rm p},\delta} = 156$~\kms.  The derived model parameters are also listed in the table.   Equation~(\ref{eq:Veff}) relates the measured effective velocity to the scintillation velocity ${\bf V}_{\rm ISS}$ which also depends on the transverse projected velocity of the Earth and the scattering screen.  Using the pulsar velocity cited above gives an estimate of the ISM screen velocity relative to the Sun ($\sim 16$~\kms), which is comparable in magnitude to other measured interstellar velocities.  

The ISM velocity analysis provides the third confirmation of the lobe-shift applied to the 1~ms astrometry, since in the alternate position $V_{\rm eff\perp}$ is reversed and could only be reconciled with the pulsar velocity if there were an implausibly high velocity ($> 200$~\kms) for the scattering screen relative to the Sun.

\begin{table}[htb]
\caption{Model Parameters for Distance and Velocity, assuming $\beta=0.353$. Note that the first five quantities are measured; the others are calculated from these assuming the pulsar distance and velocity as cited in the text.}
\begin{tabular}{lrl}
\tableline
$D_{\rm eff}$ & $1171 \pm 23$ & pc \\
$V_{\rm eff\parallel}$ & $305 \pm 3$  & \kms \\
$V_{\rm eff\perp}$ & $-145 \pm 9$ & \kms  \\
$\parallel$ Scattering axis  & $-25.2 \pm 0.5$ & deg.\ east of north \\
$\perp$ Scattering axis & $-115.2 \pm 0.5$ & deg.\ east of north \\
$D_{\rm s}$\tablenotemark{a}  &  $415 \pm 5$ & pc \\
$V_{\rm s \parallel}$\tablenotemark{b} & $-16 \pm 10$  & \kms \\
$V_{\rm s \perp}$ & $0.5 \pm 10$ & \kms  \\
$\alpha$ & $27 \pm 2$ & deg. \\
\tableline
\tablenotetext{a}{Assumes $D_{\rm p}=640$~pc; the error in $D_{\rm s}$ due to the uncertainty \\ in $D_{\rm p}$ is much larger $\sim 40$\%}
\tablenotetext{b}{Including errors from uncertainty in pulsar distance and \\proper motion}
\end{tabular}
\label{tab:params}
\end{table}

\begin{figure}[ht]
\centerline{\includegraphics[bb= 60 190 570 624,width=9cm,clip]{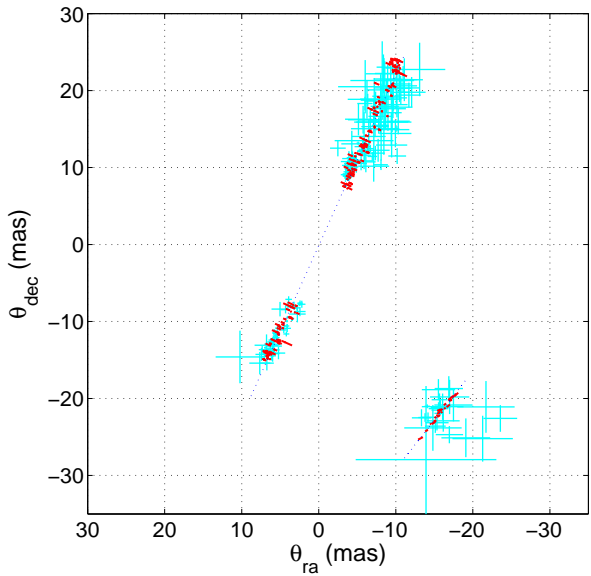}}
\caption{Cyan error bars are the VLBI astrometry of the apexes of the identified arclets superimposed from all 4 sub-bands.  Red error boxes show the much finer precision of the back-mapped astrometry, as explained in \S\ref{sec:model_map} and clearly delineate the highly linear scattering from both the main arc and the 1~ms feature (at lower right).}
\label{fig:mapped}
\end{figure}

\subsection{Estimating the Image Center}

In the discussion so far we have assumed that the astrometry is correctly centered on the emission centroid.  However, since the scattered image is highly anisotropic, the centering done by self-calibration in the primary analysis would have been more precise in the perpendicular direction than in the parallel direction.  Close examination of the initial versions of the plots as in Figures~\ref{Zeff} and \ref{fig:Vest} revealed small but significant asymmetries, which allowed us to improve the centering in the parallel direction.

In the Doppler plot of Figure~\ref{fig:Vest} a straight line fit did not pass exactly through the origin and this fitted offset in $\theta_{\parallel}$ was found to be on the order of 0.5~mas differing slightly between sub-bands.   In the delay plot of Figure~\ref{Zeff} the $\theta^2$ values at a given delay with positive Doppler frequencies were systematically shifted relative to those with negative Doppler frequencies.  We were able to minimize the $\chi^2$ for the straight line by fitting for a positional offset in the parallel direction.  We found that in each sub-band the optimum parallel offsets were consistent between the Doppler and delay estimation methods within their errors.   An average of these two offsets was consequently applied to the astrometry in each sub-band before the analysis described in the two foregoing sections.

\subsection{Astrometry Mapped Back from the Secondary Spectrum}
\label{sec:model_map}

The synthesized beam size of our VLBI array (its basic angular resolution) was about 35~mas at 327~MHz.  In the astrometry shown in Figure~\ref{AstrometryFig} we obtain positional accuracies as small as $\pm 0.5$~mas near the center of the image due to the high signal to noise ratio and much larger errors away from the center.   However we can do even better from the apex positions of the arclets in the SS using equations (\ref{eq:tau_apex}) and (\ref{eq:fDapex}), which we will refer to as ``back-mapping''.  This is possible since we used the VLBI astrometry to obtain accurate values for the model parameters $D_{\rm eff}$, ${\bf V}_{\rm eff}$, and $\alpha$. The equations give the magnitude of $\theta$ from the delay $\tau$ and the component $\theta_V$ parallel to ${\bf V}_{\rm ISS}$ from $f_{\rm D}$. There is an inherent ambiguity in such a mapping since it does not determine the sign of the angular coordinate orthogonal to the velocity.  However, we can use the VLBI image to resolve that ambiguity and show the positions of each apex mapped back into in RA, Dec in Figure~\ref{fig:mapped} overplotted for the 4 sub-bands.  The mapped astrometry points are plotted as red error boxes on a background of the error bars of the VBLI positions.   The apex points mapped back in this fashion extend along the same axis as the VLBI positions but with a narrower spread in $\theta_{\perp}$ due to their even smaller errors.

With errors $\delta\tau$ and $\delta f_{\rm D}$ in the apex positions we get positional errors  $\delta \theta \sim c\,\delta\tau/(D_{\rm eff}\theta)$ and $\delta \theta_V \sim \delta f_{\rm D}\lambda /V_{\rm eff}$.  With $\delta\tau \sim 1$~$\mu$s we obtain $\delta \theta \sim \theta_{\rm mas}^{-1}$, i.e., 0.05~mas at $\theta_{\rm mas} =20$~mas, which improves on the 0.5~mas from astrometry, particularly at large scattering angles where the S/N worsens; and with $\delta f_{\rm D}\sim 1$~mHz the resolution $\delta \theta_V \sim 0.6$~mas.  The resulting confidence region is shown by the red box which is narrow in radial $\theta$ coordinate and wide in the transverse direction.

The map emphasizes the remarkable anisotropy of the scattering disk and also shows evidence that the 1~ms feature is also highly extended along a roughly parallel direction.  This is an important aspect that will constrain any physical model for the scattering.

\begin{figure*}[ht]
\begin{tabular}{rl}
\includegraphics[bb= 24 240 532 580,width=9.3cm,clip]{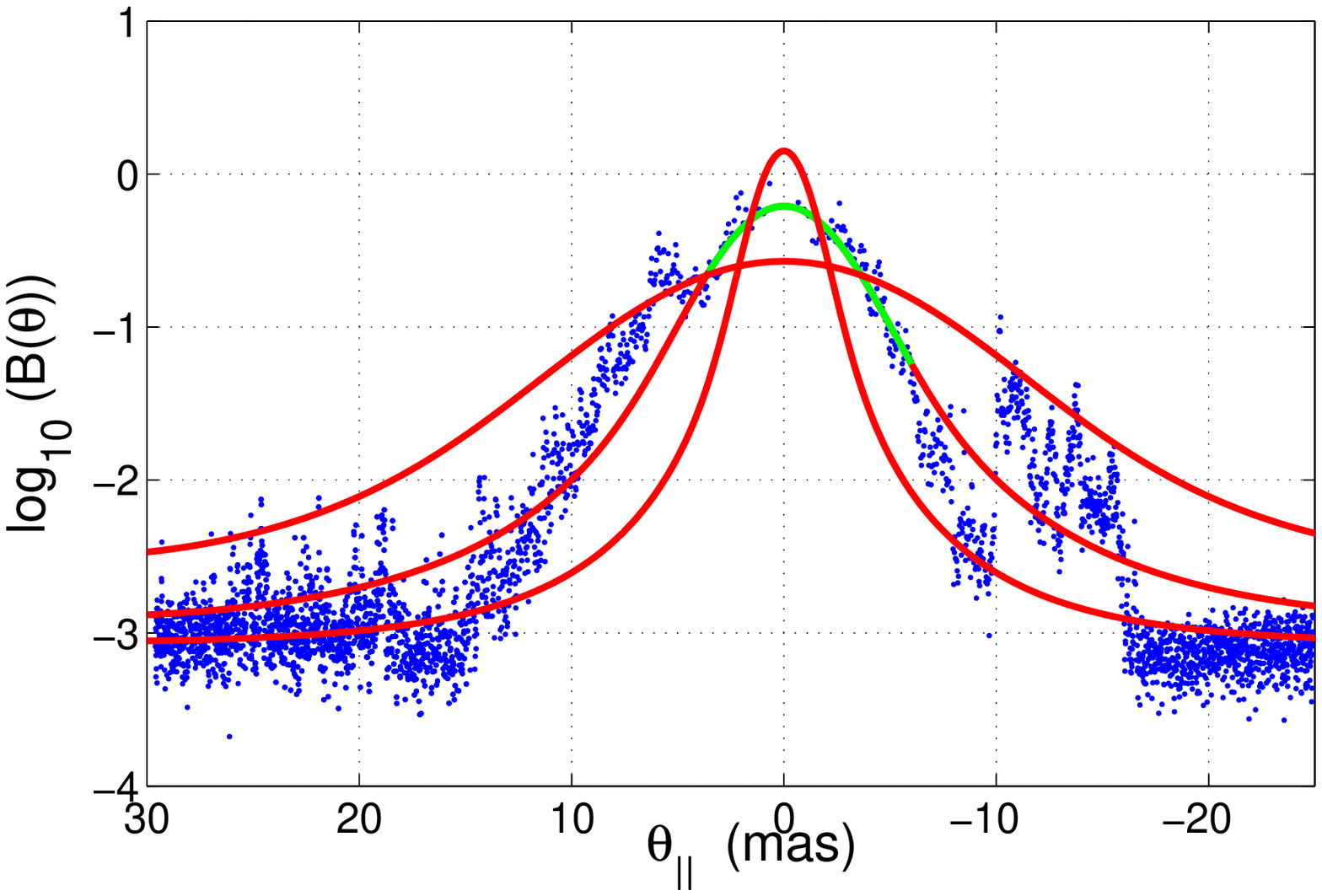} &
\includegraphics[bb= 110 240 460 580,width=6.5cm]{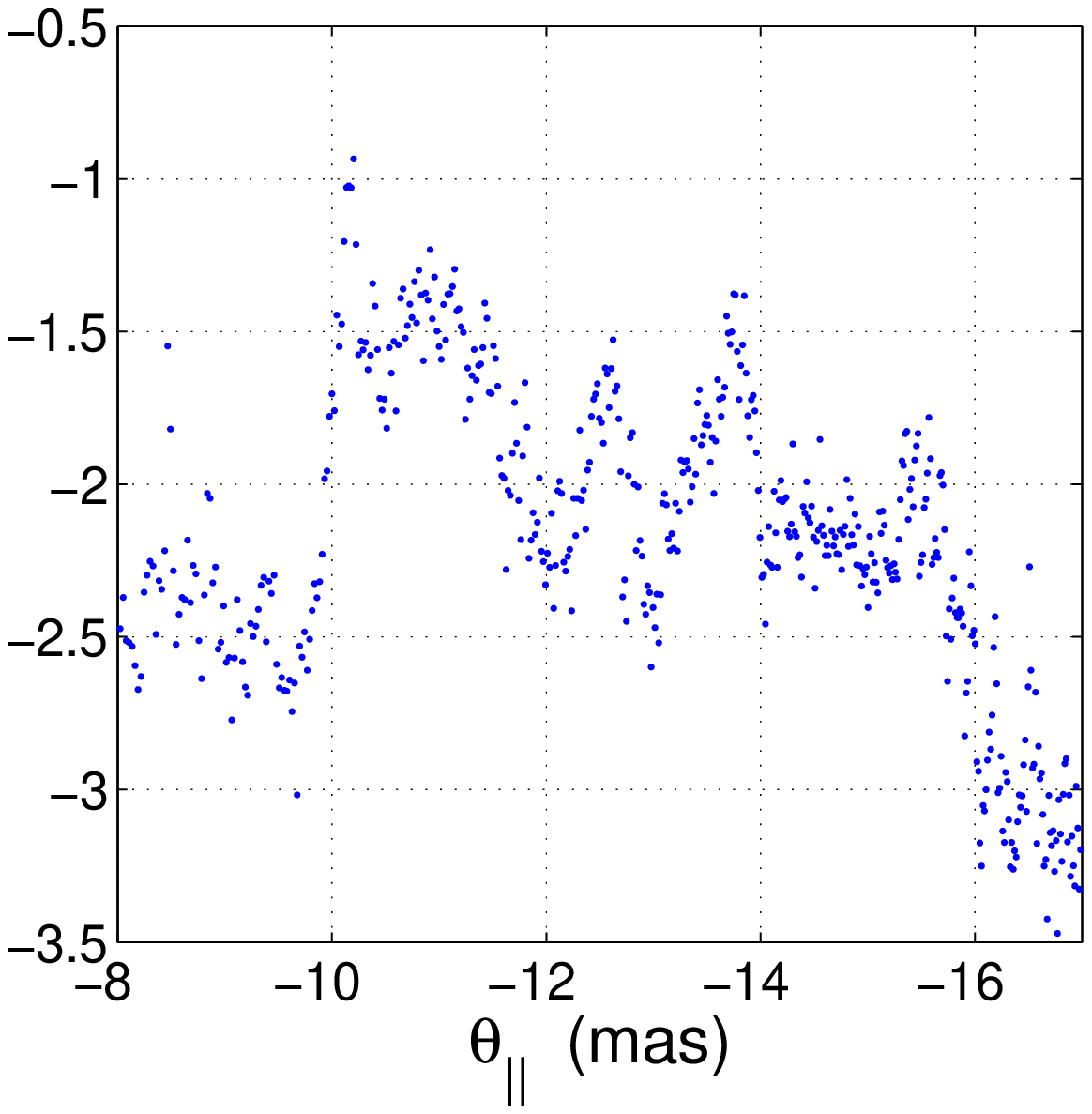} \\
\end{tabular}
\caption{\it Left:\rm Scattered brightness against $\theta_{\parallel}$ obtained for points along the main arc via the back-mapped astrometry in \S\ref{sec:model_map}, averaged from all four sub-bands (blue).  The individual peaks are as narrow as 0.1~mas as shown in the expanded view in the \it right \rm panel.  The three overplotted theoretical curves (red) are for a one-dimensional Kolmogorov model.  The middle curve was fitted to the observations over the range shown in green.  The other two curves have the same total flux density for the pulsar but are wider and narrower as discussed in the text. 
}
\label{fig:bright1d}
\end{figure*}

\subsection{Scattered Brightness Function}
\label{sec:bright}

The scattered brightness distribution can be recovered from the secondary spectrum if the intensity 
scintillations are weak, or if they are highly anisotropic \citep{Cor06}.
In this case the scintillations are 
highly anisotropic and we can make one-dimensional, strip integrated brightness distributions, $B_1(\theta_{\parallel})$, by strip integrating the two dimensional 
distribution over $\theta_{\perp}$. Then we have
\begin{eqnarray}
B_1 (\theta_1 ) B_1 (\theta_2 ) = A(\tau,f_{\rm D}) |J|
\end{eqnarray}
\noindent where $\tau = (\theta_1^2 - \theta_2^2) D_{\rm eff}/2c$, and 
$f_{\rm D} = (\theta_1 -\theta_2 ) V_{\rm eff} \cos\alpha /\lambda$
and the Jacobian $|J| = f_{\rm D} D_{\rm eff}/c$. Then by sampling $A(\tau, f_{\rm D})$ along the main arc where $\theta_2 = 0$,
we can estimate $B_1(\theta)$. This estimate is plotted in Figure~6 without the 1~ms feature, which cannot be represented on this plot since it is not on the main parabola.

If the scattering comes from homogeneous anisotropic Kolmogorov turbulence then the strip integrated brightness distribution is the Fourier transform of the one-dimensional correlation function of the electric field, $B_1(\theta_{\parallel}) = {\rm FT}$ \, $[ \exp(-(s_{\parallel} / s_0 )^{5/3})]$. We have plotted three such models over the observations in Figure~\ref{fig:bright1d}, each with a different value of $s_0$, the coherence scale of the electric field. The middle model fits the data near the origin ($s_0 \sim 10^4$~km) and the other two give a rough estimate of the range of $s_0$ necessary to match the measurements. They correspond to changing the root mean square (RMS) electron density by a factor of 2 (wider curve) or 0.5 (narrower curve). One can see in the expanded view on the right that near $\theta_{\parallel}$ = -10~mas there is a change in the RMS electron density by a factor of 4 in a very small distance. 

The finest structure has an angular scale of about 0.1~mas, which corresponds to 0.05~AU.  This plot can also be made using the VLBI astrometric $\theta_{\parallel}$. In this case the large scale structure is the same but the small scale structures seen clearly in Figure~\ref{fig:bright1d} are smoothed out by the larger error bars in the VLBI astrometry.

\subsection{Axial Ratio}

While we have clear evidence for highly anisotropic scattering, it is difficult to estimate the axial ratio $R$, which is usually defined by a contour at, say, half power in the two-dimensional scattered brightness distribution.  Estimates of the width in $\theta_{\parallel}$ can be made from the one-dimensional brightness in Figure~\ref{fig:bright1d}. However, the width at the same level in $\theta_{\perp}$ is harder to estimate, since we have no information on brightness versus $\theta_{\perp}$.  As an alternative, we can define an apparent axial ratio $R_{\rm ap}$ from the scatter of the points in Figure~\ref{AstrometryFig} independent of their brightness.  Here we define $R_{\rm ap}$ as the ratio of the RMS width in $\theta_{\perp}$ to that in $\theta_{\parallel}$. The observed scatter in $\theta_{\perp}$ will of course be broadened by the astrometric errors, so that only an upper bound can be found on $\theta_{\perp}$ due to interstellar scattering. 

\begin{figure}[ht]
\centerline{\includegraphics[bb= -10 240 580 580,width=8.5cm]{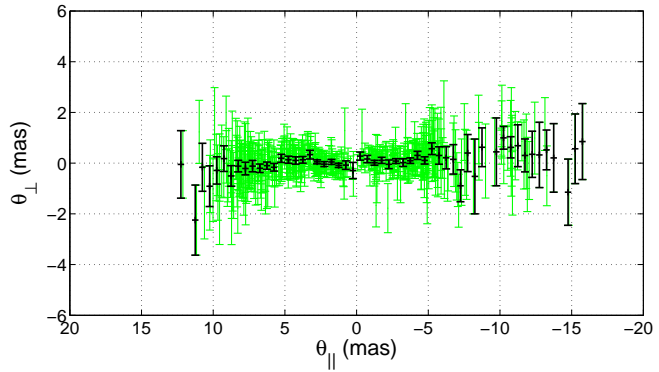}}
\caption{Astrometry referred to axes parallel and perpendicular to the main scattering axis.  Pale color superimposes scatter plots from the main arc in all four sub-bands.  The black points show the average $\theta_{\perp}$ from bins 0.5~mas wide in $\theta_{\parallel}$ with vertical bars giving the error in the average.  The bars have not been corrected for the astrometric errors. Note that only the inner portion of the primary scattering disk where high S/N points are located is included in this plot.
} 
\label{CentroidFig}
\end{figure}

We apply these ideas to Figure~\ref{CentroidFig}, which shows the astrometry from the main arc as a scatter plot of $\theta_{\perp}$ against $\theta_{\parallel}$ superimposed from all four sub-bands. The black error bars show $\theta_{\perp}$ averaged into 0.5~mas bins in $\theta_{\parallel}$ where the length of the bar is the standard deviation in each average.  Taken as a group the black points from the inner $\pm 5$~mas are consistent with a Gaussian distribution with zero mean and 0.38~mas RMS; there are no points more than $\pm 2\sigma$ from zero and their error bars, which have not been corrected for the measurement error, have a mean value of 0.3~mas.  Thus the true scatter in  $\theta_{\perp}$ is smaller than these errors, implying an intrinsic perpendicular RMS width less than $\sim 0.3$~mas.  Taking the half width at half power in $\theta_{\parallel}$ to be 3~mas from Figure~\ref{fig:bright1d} gives a lower bound on the axial ratio $R \age 10$.  Similar estimates can be made from the ratios of the parallel to the perpendicular RMS widths of the apex astrometry from both the VLBI and back-mapping methods.  These yield $R_{\rm ap} \age 27$ and $R_{\rm ap} \age 20$, respectively.

The scattering axis is neither parallel nor perpendicular to the Galactic plane. However, according to the Viriginia Tech Spectral-Line Survey\footnote{See {\tt http://www.phys.vt.edu/halpha/} for more information about this survey.} the pulsar lies $\sim 1^{\circ}$ from a $5^{\circ}$ long bright H$\alpha$ filament at a position angle within about $10^{\circ}$ of our scattering axis.  Although there is a rough agreement in position angle we have no other evidence to support an association with our 30~mas long filament.

\section{Physical Models of the Scattering}

While our main focus is the presentation of the remarkable results found from the observations, we now briefly consider what physical structures could be responsible for the scattering.  Following previous analyses of the ISS phenomena, we assume that the basic cause is scattering by random (presumably turbulent) structures in the plasma density.  The scale of such turbulence must extend down to well below the diffractive scale, $\sim \lambda/2\pi\theta_{\parallel}$. To obtain scattering angles of 25~mas requires microstructure of the order of 1000~km. However, the turbulence is inhomogeneous over scales of 0.1 to 10~AU as mentioned above and we must also account for the highly elongated shape of the main scattered image.

First we ask whether the peaks in our scattered image, due to the arclets, could be the equivalent of ``speckles''.   Discrete arclets are visible in the simulated secondary spectrum in Figure~11 of Cordes et al.\ (2006), who computed the SS for a realization of a layer with a Kolmogorov spectrum and an axial ratio 4:1 parallel to the simulated velocity.  That figure shows very many fine arclets with apexes centered on a pronounced forward parabola.  In independent realizations the arclet apexes appear at random locations, showing that they are the equivalent of speckles in a snapshot scattered image through a turbulent scattering layer. Thus they do not map to deterministic structures in the layer. 

In contrast the discrete nature of the arclets and their sparser distribution in our observation suggests deflection by discrete structures, possibly similar to those responsible for extreme scattering events (ESEs).   The two basic models proposed for ESEs have been an enhancement in $n_{\rm e}$ that causes a diverging ``lens'' \citep{Rom87} or an enhancement in plasma turbulence that causes extra scattering \citep{Fiedler87}.  The result that the angle of arrival from the arclets is essentially independent of frequency implies that the waves are deflected at a fixed transverse distance from the pulsar line-of-sight by either refraction or scattering.  This reinforces the conclusion that specific isolated structures in the line of sight are responsible.   The observed frequency independence constrains the angular size of a lens, based on the law of plasma refraction, to be a small fraction of the angle of deflection which we observe up to $\pm 20$~mas.  Similarly the size of a scattering center would have to be much smaller than this angle of deflection.

The very high apparent axial ratio we observe strongly suggests that an ordered magnetic field determines the geometry of the scattering.  The lack of any magneto-ionic signature (i.e., the exact match between the dynamic spectra in right and left circular polarizations) implies that the radio frequency phase imposed by the plasma is simply proportional to the column density of the local electron density $n_{\rm e}$.  We expect that plasma structures will be more tightly confined transverse to a magnetic field than parallel to it, i.e., the density irregularities will be field-aligned as they are in the solar corona.  Since we expect such structures to scatter preferentially in their narrow dimension, our initial idea is that the major axis of the angular scattering would be orthogonal to the mean magnetic field. Thus one scenario for the main image is a roughly parallel set of filaments (or sheets), controlled by a magnetic field at right angles to the axis of scattering and extending at least over the projected length of the image ($\sim 15$~AU).  Waves passing through a filament are scattered by 10-30~mas due to a locally enhanced column density of electrons (and presumably higher plasma turbulence) which would be seen as a single arclet.  We call this first scenario the orthogonal geometry.  The axial ratio $R$ would then correspond to the axial ratio of the micro-turbulence (i.e., at the diffractive scale).

There is however, an alternative scenario in which a dense filament of plasma is confined by the local magnetic field, which is parallel to the axis of scattering.  We call this the parallel geometry.  Such a filament must have denser knots of micro-turbulence, which are responsible for the discrete arclets.  The flux ropes seen in the solar wind are a possible plasma structure (see \citet{birn} and references therein).  The denser knots could scatter isotropically and so the filament would not have to be located exactly in front of the pulsar.  In such a case the major axis of the scattered image (about 3~mas) would be determined by diffractive scattering caused by microstructures of the order of 3000~km in scale, but the minor axis would be set by the thickness of the filament. Such scattering has been observed in the solar corona \citep{denn} and discussed in the context of interstellar scattering by \cite{Cor01}. We assume that the thickness of such a filament would be of the order of the size of a ``knot'', on the order of 0.05~AU or 0.1~mas. Thus the filament is at least 16~AU long, about 0.05~AU in diameter, but not straight. The perpendicular RMS angle of 0.4~mas would correspond to an RMS irregularity in the filament of 0.2~AU or 4 times its thickness. Because the scattering is caused by a very thin structure in this model, the electron density in the filament would have to be considerably higher than normally expected (about 5 cm$^{-3}$) to cause the observed angular scattering.

In the parallel filament model the 1~ms feature is readily explained as a separate offset filament of about the same density which is approximately parallel to the main filament.  However, in the orthogonal geometry the observer would not detect the anisotropic scattering from a concentration of filaments nearly parallel to those causing the main arc.  Thus the basic simplicity of this geometry, that assumes the scattering is so anisotropic that the angles of scattering are transverse to the long spatial axis, must be augmented by an independent process of scattering or refraction to explain the 1~ms feature.   A possible process is a localized plasma structure in that part of the sight line that refracts the scattered waves towards the observer. In this case the angular position of the 1~ms feature should have a $\lambda^{2}$ wavelength dependence, which is inconsistent with that observed as shown in \S\ref{sec:scaling}.

In either geometry the existence of many sub-AU sized discrete structures passing within 15~AU of the line of sight to our pulsar raises the more general question of their distribution in interstellar space.  The fact that the structures responsible lie at a common distance suggests that the pulsar happens to lie behind a single larger region consisting a cluster of many filaments of plasma or a single filament with very compact condensations (knots).  Further the fact that \citet{Hil05} observed similar arclets 22 months earlier suggests a region bigger than 40~AU.  
Two ideas that seem possible are
a shock with multiple nearly parallel subshocks seems and the flux rope idea already mentioned.  The number density of such regions can only be crudely constrained by the sparse statistical sampling of pulsars with similar multiple arclets.   We hope to distinguish between these two basic geometries by detailed modelling of the propagation.

\section{Conclusions}


This paper describes a novel VLBI technique resulting in a two-dimensional image of the scattering screen of pulsar B0834$+$06. The baseband data that were recorded allowed high resolution dynamic spectra to be produced. The secondary spectra produced with the two-hour dynamic spectra could allow sharply defined arclets to be identified with delays as high as 1~ms.

The scattered image was developed by astrometrically mapping points chosen from the secondary spectrum to bright points in the sky plane. These points were clustered in two clearly defined groups: a primary scattering disk which is elongated and inclined $27 \pm 2^{\circ}$ to the pulsar proper motion direction and a second, non-colinear, feature corresponding to the 1~ms feature of the secondary spectrum.  Diagnostic measurements place the two features at essentially the same distance, 65\% of the way to the pulsar. The two-dimensional distribution of points in the scattered image allows both transverse components of the effective velocity to be determined via relationships connecting the Doppler frequency with location in the image. 

The discrete feature at 1~ms delay contains about 4\% of the total received power. This feature is expected to to be visible only for a few weeks during which time its delay should drift as the pulsar moves; the impact on timing this pulsar at $\sim 327$~MHz due to such a feature is a time variable wander with magnitude $\sim 40$~$\mu$s. This should come as a caution to those aiming to perform precision pulsar timing at low frequencies on pulsars that exhibit the extreme forms of scintillation that are characteristic of B0834$+$06. Further, pulsars with sub-microsecond structure may experience apparent pulse profile evolution yielding additional complications in their timing.


We were able to estimate the effective scintillation velocity vector, which depends on a distance-weigthed sum of the velocities of the pulsar, the Earth and the sacttering plasma.  By using the published proper motion we estimated the velocity of the scattering plasma to be $16\pm 10$~\kms approximately parallel to the scattering axis.  Since the errors in this interesting result are dominated by the uncertainty in the pulsar proper motion, we have undertaken a new set of VLBI measurements to improve its precision.

The interpretation of ISS in pulsars has often assumed isotropy in the scattering.  The extremely anisotropic scattering found here would substantially alter any quantitative modelling of the plasma were it to be a common feature in other regions on the interstellar medium.  A description of the underlying plasma physics must await a resolution of the two possible geometries mentioned in the previous section, but the results and the method provide an exciting new glimpse of the ionized ISM at scales of 0.1 to 10~AU.

\acknowledgments

The Arecibo Observatory is part of the National Astronomy and Ionosphere Center, which is operated by Cornell University under a cooperative agreement with the National Science Foundation. The National Radio Astronomy Observatory is a facility of the National Science Foundation operated under cooperative agreement by Associated Universities, Inc.  This work has been supported by the Australian Federal Government's Major National Research Facilities program.   The Virginia Tech Spectral-Line Survey (VTSS) is supported by the National Science Foundation.  WFB thanks Shri Kulkarni for his hospitality during a visit to Caltech in the course of this work.  Thanks go to Scott Ransom for helping with the pulsar timing that enabled pulsar gating.  BJR and WAC are grateful for support of scintillation research at UCSD under the grant AST 0507713 from the NSF.  The authors thank an anonymous referee for providing input that has made the complicated details of this paper more comprehendable.

\end{document}